\documentclass[11pt,a4paper]{article}
\usepackage{amssymb,epsf,epsfig,amsmath,color}
\usepackage{multirow}
\usepackage{accents}

\newcommand{\lsim}{
\mathrel{\hbox{\rlap{\hbox{\lower4pt\hbox{$\sim$}}}\hbox{$<$}}}}
\newcommand{\gsim}{
\mathrel{\hbox{\rlap{\hbox{\lower4pt\hbox{$\sim$}}}\hbox{$>$}}}}

\def\D0{D\O }

\def\theabstract{CP violation in $B^0_s$--$\bar B^0_s$ oscillations is expected at the $10^{-5}$ level in the
Standard Model but could be enhanced by New Physics. Using $B^0_s\to D_s^-\ell^+\nu_\ell$
decays, LHCb has recently reported the new result $(0.39\pm0.33)\times10^{-2}$ of the 
corresponding observable $a^s_{\rm sl}$. We point out that other current $B$ decay data imply 
$a^s_{\rm sl} = (0.004 \pm 0.075)\times10^{-2}$. In view of this strong constraint, we propose to use 
$B^0_s\to D_s^-\ell^+\nu_\ell$ and similar flavor-specific decays 
as a new tool to determine both the production asymmetry between $B^0_s$ and $\bar B^0_s$ 
mesons, and the CP asymmetry in the subsequent $D^\pm_s$ decays. The former serves 
as input for analyses of CP violation in $B^0_s$ channels, with significant room for 
improvement, while the latter offers an exciting laboratory for New Physics.}
\begin{document}
\begin{titlepage}
\vspace*{-0.7truecm}
\begin{flushright}
Nikhef-2016-029\\
SI-HEP-2016-17 \\
QFET-2016-11
\end{flushright}

\vspace{0.6truecm}

\begin{center}
\boldmath
{\Large{\bf $B^0_s$--$\bar B^0_s$ Oscillations as a New Tool to \\
Explore CP Violation in $D_s^\pm$ Decays}
}
\unboldmath
\end{center}

\vspace{1.2truecm}

\begin{center}
{\bf Robert Fleischer\,${}^{a,b}$ and  K. Keri Vos\,${}^{a,c,d}$}

\vspace{0.5truecm}

${}^a${\sl Nikhef, Science Park 105, NL-1098 XG Amsterdam, Netherlands}

${}^b${\sl  Department of Physics and Astronomy, Vrije Universiteit Amsterdam,\\
NL-1081 HV Amsterdam, Netherlands}

${}^c${\sl Van Swinderen Institute for Particle Physics and Gravity, University of Groningen,\\
NL-9747 AG Groningen, Netherlands}

${}^d${\sl Theoretische Physik 1, Naturwissenschaftlich-Technische Fakult\"at, \\
Universit\"at Siegen, D-57068 Siegen, Germany}

\end{center}

\vspace*{1.7cm}

\begin{center}
\large{\bf Abstract}\\

\vspace*{0.6truecm}

\begin{tabular}{p{13.5truecm}}
\theabstract
\end{tabular}

\end{center}

\end{titlepage}

\thispagestyle{empty}
\vbox{}
\newpage

\setcounter{page}{1}

\vspace*{-2.5truecm}

\section{Introduction}
Studies of CP violation provide interesting tests of the Standard Model (SM) of particle physics, where 
decays of neutral $B^0_s$ mesons play a key role at the Large Hadron Collider (LHC) \cite{BFS-rev}. 
These particles show $B^0_s$--$\bar B^0_s$ mixing, which in the SM is generated through quantum 
fluctuations. New Physics (NP) may affect $B^0_s$--$\bar B^0_s$ mixing through contributions 
at the tree level, mediated, for instance, through $Z'$ bosons, or through new heavy particles running 
in the loop diagrams \cite{BF}. 

CP violation in $B^0_s$--$\bar B^0_s$ oscillations is described by an observable $a^s_{\rm sl}$ and
is vanishingly small in the SM \cite{ABL}:
\begin{equation}\label{asl_SM}
a^s_{\rm sl}|_{\rm SM}=(2.22\pm0.27)\times10^{-5},
\end{equation}
but could be enhanced by NP. However, in recent years, a wealth of experimental information on 
$B^0_s$--$\bar B^0_s$ mixing and CP violation in $B$-meson decays was obtained, in 
accordance with the SM. In view of this progress, the question 
arises how much space for NP effects in $a^s_{\rm sl}$ is actually left by the data. This important issue, 
which can in fact be raised for many flavor-physics observables, is the key motivation of the 
following discussion.

The observable $a^s_{\rm sl}$ can be measured through semileptonic $B^0_s\to D_s^-\ell^+\nu_\ell$ 
and $\bar B^0_s\to D_s^+\ell^-\bar\nu_\ell$ decays \cite{DFN}. In the SM, such transitions are 
flavor-specific: 
\begin{equation}\label{flav-spec}
A(B^0_s\to D_s^+\ell^-\bar\nu_\ell)=A(\bar B^0_s\to D_s^-\ell^+\nu_\ell)=0,
\end{equation}
such that the ``wrong-sign" decays $B^0_s\to D_s^+\ell^-\bar\nu_\ell$ and 
$\bar B^0_s\to D_s^-\ell^+\nu_\ell$ can only occur through $B^0_s$--$\bar B^0_s$ mixing. 
The LHCb collaboration has recently reported the world's best 
measurement for $a^s_{\rm sl}$ \cite{LHCb-asl}:
\begin{equation}\label{LHCb}
a^s_{\rm sl}=\left[0.39\pm0.26{\rm (stat)}\pm0.20 {\rm (syst)}\right]\times10^{-2},
\end{equation}
which agrees with the SM prediction (\ref{asl_SM}). The average of the previous results 
is given as follows \cite{HFAG}:
\begin{equation}\label{HFAG-av}
a^s_{\rm sl}=-(0.48\pm0.48)\times10^{-2}.
\end{equation}
Here the \D0 result  $a^s_{\rm sl}=-(1.33\pm0.58)\times10^{-2}$ \cite{D0-dimuon},
which differs from the SM at the $3\sigma$ level and led to attention in the community
(see, e.g., \cite{BFS-rev,Lenz-12}), was not included.

Using measurements of $B^0_s$--$\bar B^0_s$ mixing and CP violation in $B$ 
decays caused by $b\to c \bar c s$ processes, we show that $a^s_{\rm sl}$ is constrained -- in
a model-independent way -- at the $10^{-4}$ level. 

In view of this strong constraint, we propose a new method to utilize flavor-specific $B^0_s$ decays. It allows 
the determination of the $B^0_s$--$\bar B^0_s$ production asymmetry and opens a new avenue to 
explore CP violation in $D_s^\pm$ decays, which is tiny in the SM but 
may be enhanced through NP effects. The impact of possible CP violation in $D_s^\pm$ decays has 
not been included in the LHCb result (\ref{LHCb}). We shall take this effect into account in our analysis 
to show the sensitivity of the new strategy.

\boldmath
\section{A Closer Look at $a_{\rm sl}^s$}\label{sec:aSL}
\unboldmath
The observable $a_{\rm sl}^s$ takes the following form \cite{DFN}:
\begin{equation}\label{aSL-DGM}
a_{\rm sl}^s=\left|\frac{\Gamma_{12}^{(s)}}{M_{12}^{(s)}}\right|\sin\tilde\phi_s =\left(\frac{\Delta 
\Gamma_s}{\Delta M_s}\right)\tan(\tilde\phi_s),
\end{equation}
where $\Gamma_{12}^{(s)}$ and $M_{12}^{(s)}$ are the off-diagonal elements of the
decay and mass matrices describing $B^0_s$--$\bar B^0_s$ mixing, $\Delta M_s$ and 
$\Delta\Gamma_s$ are the mass and decay width differences of the $B^0_s$ mass eigenstates, 
respectively, and 
\begin{equation}\label{phi_s-tilde-def}
\tilde \phi_s=\arg(-M_{12}^{(s)}/\Gamma_{12}^{(s)})
\end{equation}
is a CP-violating phase difference. As $M_{12}^{(s)}$ is governed by short-distance contributions, 
NP may have a significant impact. On the other hand, the matrix element 
\begin{equation}\label{Gamma12-def}
\Gamma_{12}^{(s)}=\sum_f{\cal N}_f\langle B_s^0|f\rangle\langle f|\bar B^0_s\rangle,
\end{equation}
where ${\cal N}_f$ is a phase-space factor \cite{DFN}, is dominated by tree decays caused by 
$b\to c\bar c s$ processes, which are favoured by the Cabibbo--Kobayashi--Maskawa 
(CKM) matrix, and is hence expected to be insensitive to NP contributions \cite{DFN,Lenz-12, Bad07}. 
Detailed theoretical studies of $\Gamma_{12}^{(s)}$ were performed in \cite{BGP} and \cite{BH,BHLPT}, 
where the latter analyses were motivated in particular by the \D0 result \cite{D0-dimuon}. These studies 
found smallish room for NP effects in $\Gamma_{12}^{(s)}$, also through poorly constrained 
$(\bar{s} b$)$(\bar{\tau} \tau)$ operators. 

The Particle Data Group (PDG) \cite{PDG} gives the averages 
\begin{equation}\label{DGs-xs}
\frac{\Delta \Gamma_s}{\Gamma_s}=0.124\pm0.011, 
\quad
x_s\equiv\frac{\Delta M_s}{\Gamma_s}=26.81\pm0.10,
\end{equation}
where $1/\Gamma_s=(1.510\pm0.005)\times10^{-12}\,\mbox{s}$ is the $B^0_s$ lifetime. The 
experimental results for $\Delta\Gamma_s$ and $\Delta\Gamma_s/\Delta M_s$ are consistent with 
the SM predictions although the theoretical uncertainties are still at the 20\% level \cite{ABL}. For
a discussion of NP effects on $\Delta\Gamma_s$ in multi-Higgs and left-right-symmetric models,
see \cite{BGP}. 

In the following discussion, we do not have to rely on calculations of 
$\Delta \Gamma_s$ in the SM or NP models but shall rather utilize the measured value 
of this quantity. Inserting the experimental results in (\ref{DGs-xs}) into (\ref{aSL-DGM}) yields
\begin{equation}\label{asl-1}
a_{\rm sl}^s=\left[(0.46 \pm0.04) \times 10^{-2}\right] \times \tan(\tilde\phi_s).
\end{equation}
It is interesting to note that the numerical pre-factor pushes this observable already into the 
regime of (\ref{LHCb}). 

Let us now exploit measurements of CP violation in $B^0_s$ decays. Using (\ref{phi_s-tilde-def}) and
(\ref{Gamma12-def}), and writing the $\bar B^0_s\to f$ decay amplitudes for a final state 
$f= J/\psi \phi, D_s^-D_s^+, ...$\ caused by $b \to c \bar c s$ processes in the following 
general way
\begin{equation}\label{ampl-f}
\bar A_f=|\bar A_f| e^{i[{\rm arg}(V_{cb}V_{cs}^*)+\bar\psi_f]},
\end{equation}
we obtain
\begin{equation}\label{asl_det}
a_{\rm sl}^s=\left(\frac{\Delta \Gamma_s}{\Delta M_s}\right)\tan(\langle \phi_s\rangle + \Delta\Psi).
\end{equation}
The phase $\langle \phi_s\rangle$ is the average of 
\begin{equation}
\phi_f = \phi_s^{\rm SM}+\phi_s^{\rm NP}+\Delta \psi_f,
\end{equation}
where $\phi_s^{\rm SM}=-(2.1\pm0.1)^\circ$ \cite{ABL}, $\phi_s^{\rm NP}$ originates from 
CP-violating NP contributions to $M_{12}^{(s)}$, and $\Delta \psi_f\equiv\psi_f-\bar\psi_f$, where 
the signs of the CP-violating phases entering $\psi_f$ are reversed with respect to $\bar\psi_f$. 
Measurements of mixing-induced and direct CP asymmetries allow the extraction of $\phi_f$ \cite{FFM}:
\begin{equation}
\frac{{\cal A}_{\rm CP}^{\rm mix}(B_s\to f)}{\sqrt{1-{\cal A}_{\rm CP}^{\rm dir}(B_s\to f)^2}}=
\eta_f\sin\phi_f,
\end{equation}
where $\eta_f$ is the CP eigenvalue of the final state. In order to determine $\langle \phi_s\rangle$
from the experimental data, we use
\begin{equation}
\langle \phi_s\rangle = \frac{\sum_f\phi_f/\sigma_f^2}{\sum_f 1/\sigma_f^2}
\pm\biggl(\sum_f 1/\sigma_f^2\biggr)^{-1/2},
\end{equation}
where the measured values take the form $\phi_f\pm\sigma_f$ \cite{PDG}. 
The process dependence of $\phi_f$ enters through
\begin{equation}
\Delta\psi_f=\Delta\psi_f^{\rm SM}+\Delta\psi_f^{\rm NP},
\end{equation}
where the SM piece $\Delta\psi_f^{\rm SM}$ is caused by doubly Cabibbo-suppressed penguin 
topologies, while $\Delta\psi_f^{\rm NP}$ originates from NP contributions to $b\to c \bar c s$ 
processes. Using data for control channels, $\Delta\psi_f^{\rm SM}$ is constrained to be at 
most ${\cal O}(0.5^\circ)$ for $B^0_s\to J/\psi \phi$ \cite{DeBrFl} and $-(1.7^{+1.7}_{-1.4})^\circ$ for 
$B^0_s\to D_s^-D_s^+$ \cite{BDD}.  The phase shift $\Delta\Psi$ is given by
\begin{equation}\label{DeltaPsi}
\Delta\Psi = \mbox{arg}\Biggl[\sum_f \eta_f w_f e^{i(\phi_f-\langle \phi_s\rangle)}\Biggr]
\end{equation}
with
\begin{equation}\label{wf-def}
w_f= \Gamma(B^0_s\to f)\sqrt{\frac{1-{\cal A}_{\rm CP}^{\rm dir}(B_s\to f)}{1+
{\cal A}_{\rm CP}^{\rm dir}(B_s\to f)}},
\end{equation}
where $\Gamma(B^0_s\to f)$ is the rate of the corresponding decay. As discussed in \cite{DFN},
any final state $|f\rangle$ can be decomposed in its CP-even and CP-odd components 
$|f_{\rm CP+}\rangle$ and $|f_{\rm CP-}\rangle$, respectively, and the sum actually 
runs only over these states, i.e.\ interference terms involving $\langle f_{\rm CP+}|B^0_s\rangle$ 
and $\langle f_{\rm CP-}|B^0_s\rangle$ drop out in the sum.

Decays of $\bar B^0_s$ mesons caused by $b\to c \bar u s, u \bar c s$ processes give sub-leading
contributions to (\ref{Gamma12-def}), with the ratio of the corresponding CKM factors given by
$\lambda^2 R_be^{i\gamma}\approx 0.02 \times e^{i 70^\circ}$, where $\lambda\approx0.2$ is the
Wolfenstein parameter, $R_b\approx0.4$ is one side of the Unitarity Triangle, and $\gamma$ 
one of its angles. The impact of these contributions
on the phase in (\ref{phi_s-tilde-def}) is hence of ${\cal O}(1^\circ)$. The difference  
$\tilde\phi_s^{\rm SM}-\phi_s^{\rm SM}$ actually probes these terms \cite{ABL}, 
and the calculated SM value at the $2^\circ$ level agrees with our general expectation. 
Assuming CP-violating NP contributions to the sub-leading tree $\bar B^0_s$ decays at the $10\%$ 
level gives a tiny phase shift of $\tilde\phi_s$ at the ${\cal O}(0.1^\circ)$ level, which is 
irrelevant for our considerations.

\begin{figure}[tbp] 
   \centering
  \includegraphics[width=2.6in]{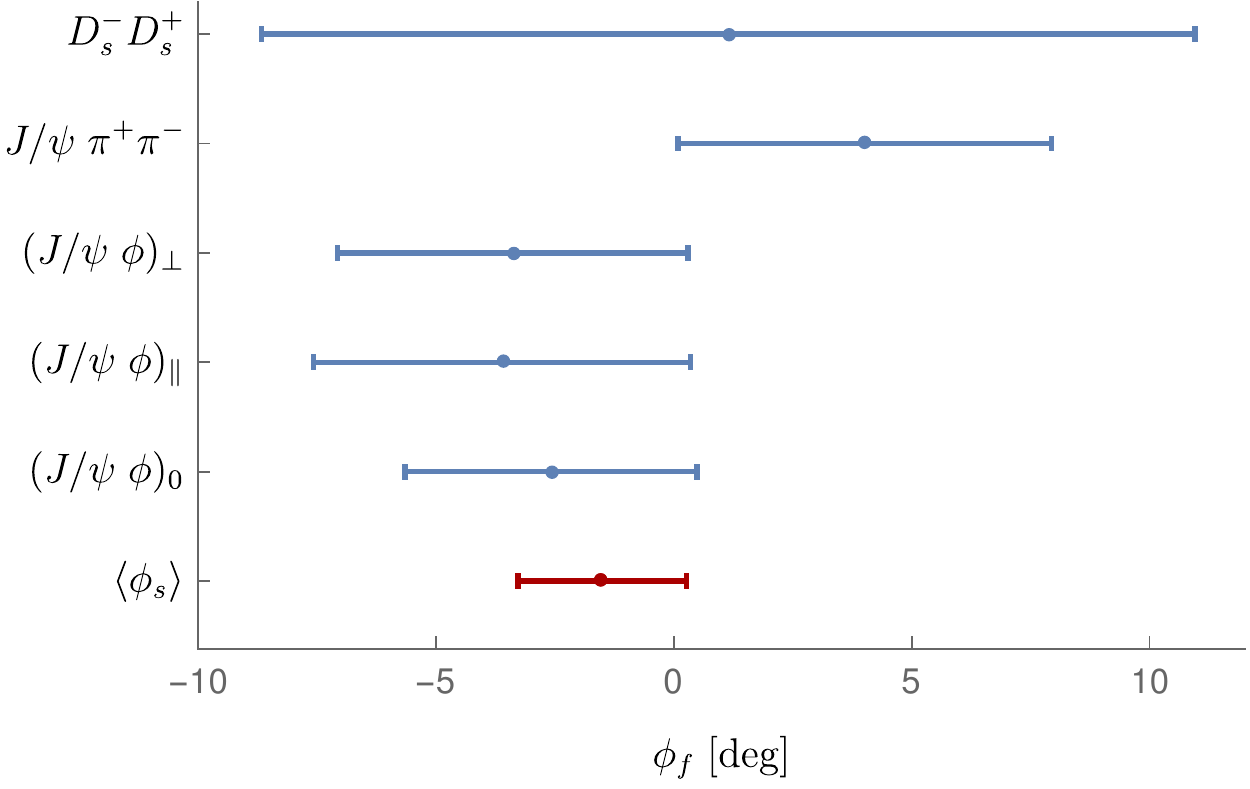} 
  \vspace*{-0.4truecm}
   \caption{Compilation of the available measurements of $\phi_f$ for various $B^0_s\to f$ decays 
   originating from $\bar b \to \bar c c \bar s$ processes.}\label{fig:1}
\end{figure}

\begin{figure}[tbp] 
   \centering
  \includegraphics[width=2.6in]{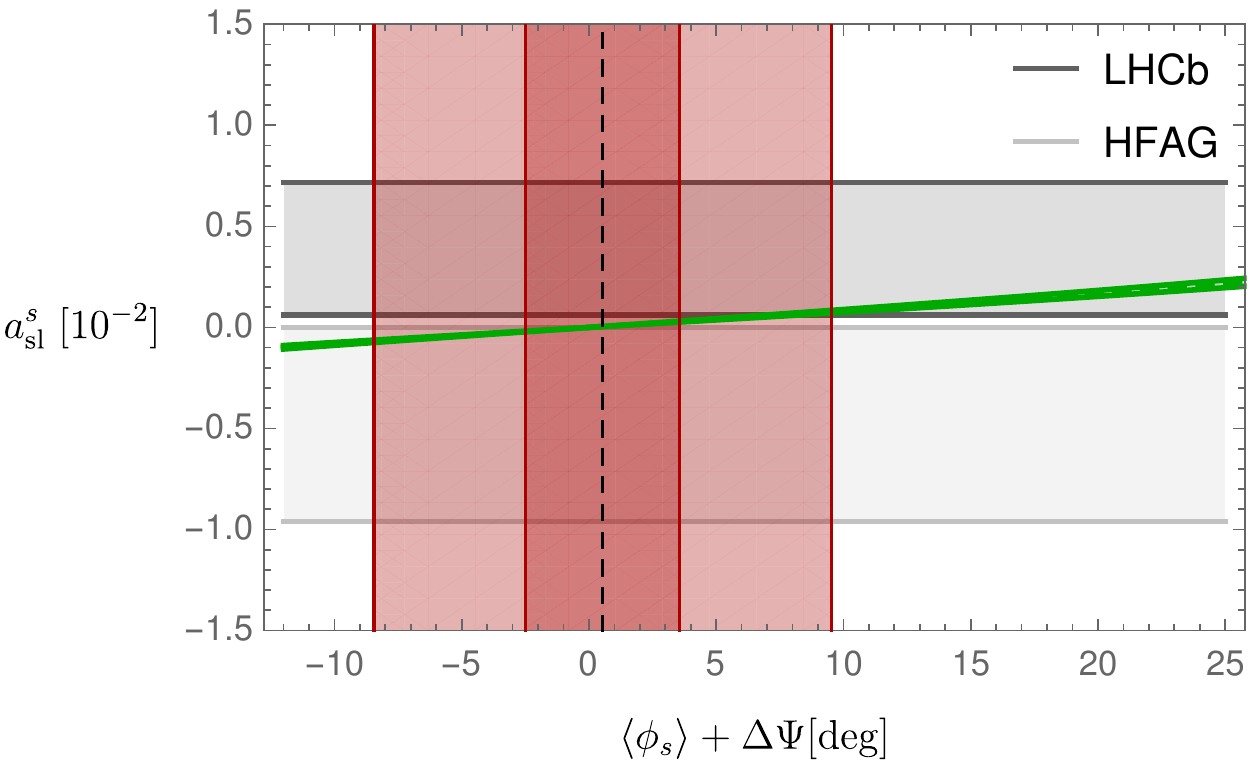} 
  \vspace*{-0.4truecm}
   \caption{Dependence of $a_{\rm sl}^s$ on $\langle\phi_s\rangle+\Delta\Psi$ 
   following from (\ref{asl-1}) and (\ref{asl_det}). The vertical bands correspond to the experimental
    range in (\ref{asl-num-1}) and (\ref{asl-num-2}), 
    while the horizontal bands show the LHCb and HFAG 
    results in (\ref{LHCb}) and (\ref{HFAG-av}), respectively.}\label{fig:2}
\end{figure}

Expressions (\ref{ampl-f}--\ref{wf-def}) are general and do not rely on specific assumptions for NP
contributions to the $b\to c\bar c s$ transitions. The remarkable feature is that 
{\it experimental data} for CP asymmetries and decay rates allow us to determine the 
phase entering (\ref{asl_det}), thereby pinning down the observable $a_{\rm sl}^s$ in a 
model-independent way. 

Making the plausible assumption that NP enters only through $B^0_s$--$\bar B^0_s$ mixing \cite{DFN,GNW}, 
it was found that the measurement of CP violation in $B^0_s\to J/\psi \phi$ rules out a large enhancement 
of $a_{\rm sl}^s$  \cite{LPP,CKMfitter,UTfit}. We can now go beyond this finding by including possible 
NP contributions to $\Gamma_{12}^{(s)}$ through further data on CP violation. In Fig.~\ref{fig:1}, 
we collect the various LHCb results for $\phi_f$ that are currently available.

Concerning the $B^0_s\to J/\psi \phi$ decay, it is crucial to have the pioneering measurements of the 
different CP-even ($0$, $\parallel$) and CP-odd ($\perp$) 
final-state configurations \cite{LHCb-Bspsiphi}. The LHCb analysis of $B^0_s\to J/\psi \pi^+\pi^-$ 
\cite{LHCb-Bspsipipi} is actually largely dominated by the CP-odd $B^0_s\to J/\psi f_0(980)$ 
contribution \cite{SZ,FKR}. These measurements do not reveal a process dependence within 
the uncertainties and are consistent with the SM pattern of tiny values of $\phi_f$. Since an 
accidental cancellation between $\phi_s^{\rm NP}$ and the $\Delta\psi_f^{\rm NP}$ is not plausible, 
we conclude that these NP phases are all small. This picture is also supported by data for 
$B^0_d\to J/\psi K^0$ modes which do not show any sign of direct CP violation at the 1\% level; 
for $B^\pm\to J/\psi K^\pm$ decays, such effects are even constrained to vanish at the 0.3\% 
level \cite{PDG}. Should there be an accidental cancellation between $\phi_s^{\rm NP}$ and the 
$\Delta\psi_f^{\rm NP}$ for some subset of final states, it would not affect our analysis of $a_{\rm sl}^s$
as the general expressions in (\ref{ampl-f}--\ref{wf-def}) do not rely on specific assumptions for NP
affecting  the $b\to c\bar c s$ processes and are also valid in this situation.

In the case of $B^0_s$ decays with open charm, we have only a first study of CP violation in 
$B^0_s\to D_s^-D_s^+$ \cite{LHCb-BsDsDs}, which has a significant uncertainty. However, we may 
probe NP also through $B^+\to \bar D^0D_s^+$. The Belle collaboration has measured the 
direct CP asymmetry of this channel as $(0.5\pm1.5)\%$ \cite{Belle-dir}, which should be 
compared with ${\cal A}_{\rm CP}^{\rm dir}(B_s\to D_s^-D_s^+)=(9.0\pm 20)\%$ 
and does not indicate any deviation from the SM with high precision. Assuming a NP contribution
with sizable CP-violating and CP-conserving phase differences, the $B^+\to \bar D^0D_s^+$
result corresponds to $\Delta\psi_{D_s^-D_s^+}^{\rm NP}$ in the few degree regime, 
in full agreement with the data for decays of the kind $B^0_s\to J/\psi \phi$ discussed in the previous 
paragraph.

The average of the measurements in Fig.~\ref{fig:1} is given by 
$\langle\phi_s\rangle = -(1.5 \pm 1.8)^\circ$. Applying (\ref{DeltaPsi}) to the corresponding 
final states gives $\Delta\Psi=(2.1 \pm 9.0)^\circ$, which yields
\begin{equation}\label{asl-num-1}
a_{\rm sl}^s = (0.004\pm 0.075)\times10^{-2}.
\end{equation}
This analysis can be refined through improved measurements of CP violation in the 
various channels, in particular for $B^0_s\to D_s^-D_s^+$ and $B^0_s\to D_s^{*-}D_s^{*+}$ modes, 
where in the latter case -- in analogy to $B^0_s\to J/\psi \phi$ -- polarization-dependent measurements 
are required \cite{DDF}. Analyses of CP violation in $B^+\to \bar D^{(*)0}D_s^{(*)+}$ and 
$B^0_d\to D_d^{(*)-}D_s^{(*)+}$ will further complement the picture. Let us consider a future scenario
where we reduce the error of the current measurement of $\phi_{D_s^-D_s^+}$ by a factor 
of three as an experimental benchmark, which would match the current experimental precision for 
$B^0_s\to J/\psi \phi$, resulting in $\langle\phi_s\rangle = -(1.0 \pm 1.6)^\circ$, 
$\Delta\Psi=(1.5 \pm 2.8)^\circ$ and
\begin{equation}\label{asl-num-2}
a_{\rm sl}^s =(0.004 \pm 0.024)\times 10^{-2}. 
\end{equation}

In Fig.~\ref{fig:2}, we illustrate the situation for $a^s_{\rm sl}$, taking also the measurements of 
$\Delta\Gamma_s$ and $\Delta M_s$ into account. The CP violation measurements lead to 
a dramatic further suppression of $a^s_{\rm sl}$ with respect to the numerical factor in (\ref{asl-1}). 
While $a_{\rm sl}^s$ could still be enhanced with respect to the SM prediction 
(\ref{asl_SM}), it is on the other hand already constrained to be at least a factor of four
smaller than the uncertainty of the LHCb measurement (\ref{LHCb}); the range in (\ref{asl-num-2}) 
puts an even stronger constraint. The comparison with the LHCb 
and HFAG bands shows impressively that $a_{\rm sl}^s$ is strongly constrained by currently available
data despite the possible impact of NP contributions. This finding answers the key question about the 
space left for NP in this observable. Nevertheless it would be interesting to confront this picture with 
more precise measurements of the  $a_{\rm sl}^s$ observable, and new ideas on the experimental side
could result in more progress than currently foreseen. 

\section{The New Strategy}\label{sec:rate-asym}
In view of the constraints on $a_{\rm sl}^s$ derived in the previous section, flavor-specific $B^0_s$ 
decays offer an interesting new playground. Assuming (\ref{flav-spec}), 
as is usually done in the literature, $B^0_s\to D_s^-\ell^+\nu_\ell$ and 
$\bar B^0_s\to D_s^+\ell^-\bar\nu_\ell$ are flavor-specific transitions. Interestingly, these relations 
have not yet been tested by experiment. In the SM, they receive corrections from processes of 
higher order in electroweak interactions, which are extremely small. But as NP may, in principle, 
have some impact, we give the most general expressions for the relevant observables, allowing us 
to search for violations of (\ref{flav-spec}). To simplify the discussion, we keep only leading-order 
terms of small parameters.

Following \cite{DFN}, we introduce
\begin{equation}\label{ampl-rel}
\lambda = - e^{-i\phi_{\rm M}^{(s)}}\left[\frac{A(\bar B^0_s\to 
D_s^-\ell^+\nu_\ell)}{A(B^0_s\to D_s^-\ell^+\nu_\ell)}\right],
\end{equation}
where $\phi_{\rm M}^{(s)}$ is the CP-violating phase associated with 
$B^0_s$--$\bar B^0_s$ mixing, $\bar\lambda$ involves the CP-conjugate decays, and
\begin{equation}\label{ACP-diff}
\Delta{\cal A}_{\rm CP}^{\rm mix} = -2\,{\rm Im}\,(\lambda-\bar\lambda), \quad
\Delta{\cal A}_{\Delta\Gamma} = -2\,{\rm Re}\,(\lambda-\bar\lambda) \ .
\end{equation}
The formalism is analogous to $B^0_s\to D_s^+K^-$, $B^0_d\to D^+\pi^-$ decays and is discussed
in detail in Ref.~\cite{RF-BsDsK}. There it is also shown explicitly that the combination of the 
convention-dependent mixing phase $e^{-i\phi_{\rm M}^{(s)}}$ with the amplitude ratio in 
(\ref{ampl-rel}) actually results in convention-independent observables $\lambda$ and $\bar\lambda$.

If (\ref{flav-spec}) holds, $\lambda$, $\bar\lambda$ and the observables in (\ref{ACP-diff}) vanish. 
It is useful to define the time-dependent functions
\begin{equation}
F_\pm(t)\equiv
\frac{{\Delta\cal A}_{\rm CP}^{\rm mix}\sin\Delta M_st \pm
\Delta{\cal A}_{\Delta\Gamma}\sinh \Delta\Gamma_st/2}{2(\cos \Delta M_st
\pm\cosh \Delta\Gamma_st/2)} .
\end{equation}

Let us now consider the following ``wrong-sign" asymmetry for the time-dependent decay rates:
\begin{equation}\label{aW-1}
a_{\rm WS}\equiv \frac{\Gamma(\bar B^0_s(t)\to D_s^-\ell^+\nu_\ell)-
\Gamma(B^0_s(t)\to D_s^+\ell^-\bar\nu_\ell)}{\Gamma(\bar B^0_s(t)\to D_s^-\ell^+\nu_\ell)+
\Gamma(B^0_s(t)\to D_s^+\ell^-\bar\nu_\ell)}.
\end{equation}
For $t>0$, and taking into account that $a_{\rm WS}$ is obtained experimentally by 
measuring the number of decay events, it takes the form
\begin{equation}\label{aWS-expr}
a_{\rm WS}=A_{\rm P}(B_s)+a_{\rm CP}(\ell^+\nu_\ell;f_{D_s}) +a_{\rm sl}^s + F_-(t),
\end{equation}
where the time dependence allows us to test (\ref{flav-spec}). If we assume (\ref{flav-spec}), 
the time dependence cancels and the time-dependent rates in (\ref{aW-1}) can be replaced by their 
time-integrated counterparts. The production asymmetry 
\begin{equation}
A_{\rm P}(B_s)\equiv \frac{\sigma(\bar B^0_s)-\sigma(B^0_s)}{\sigma(\bar B^0_s)+\sigma(B^0_s)}, 
\end{equation}
where $\sigma(\bar B^0_s)$ and $\sigma(B^0_s)$ 
denote production cross sections,  enters studies of CP 
violation and is a non-perturbative, hadronic quantity which is characteristic for the environment where the 
mesons are produced. The LHCb measurement $A_{\rm P}(B_s)=(1.09\pm2.61\pm0.66)\%$ 
leaves a lot of room for improvement \cite{LHCb-prod-asym}. The CP asymmetry 
\begin{equation}\label{acp-def}
a_{\rm CP}(\ell^+\nu_\ell;f_{D_s})\equiv \frac{\Gamma(B^0_s\to D_s^-[\to f_{D_s}]\ell^+\nu_\ell)-
\Gamma(\bar B^0_s\to D_s^+[\to \bar f_{D_s}]\ell^-\bar\nu_\ell)}{\Gamma(B^0_s\to D_s^-[\to f_{D_s}]
\ell^+\nu_\ell)+\Gamma(\bar B^0_s\to D_s^+[\to \bar f_{D_s}]\ell^-\bar\nu_\ell)}
\end{equation}
of the time-independent decay rates (i.e.\ at $t=0$), where $f_{D_s}$ ($\bar f_{D_s}$) is the final state 
of the subsequent $D_s^-$ ($D_s^+$) decay, may reveal new sources of CP violation and will be 
discussed in detail in Section~\ref{sec:directCPV}. 

The observable $a_{\rm WS}$ can be complemented with the ``right-sign" lepton asymmetry
\begin{equation}
a_{\rm RS}\equiv \frac{\Gamma(\bar B^0_s(t)\to D_s^+\ell^-\bar\nu_\ell)-
\Gamma(B^0_s(t)\to D_s^-\ell^+\nu_\ell)}{\Gamma(\bar B^0_s(t)\to D_s^+\ell^-\bar\nu_\ell)+
\Gamma(B^0_s(t)\to D_s^-\ell^+\nu_\ell)},
\end{equation}
where the final states can be accessed directly, i.e.\ without $B^0_s$--$\bar B^0_s$ oscillations
or a violation of (\ref{flav-spec}). It takes the following form:
\begin{equation}
a_{\rm RS}=A_{\rm P}(B_s) - a_{\rm CP}(\ell^+\nu_\ell;f_{D_s}) - F_+(t),
\end{equation}
where the time-dependent function allows us again to probe (\ref{flav-spec}). Assuming the 
relations given there, $a_{\rm RS}$ can be extracted from the tagged, time-integrated rates.

As we have seen in Section~\ref{sec:aSL}, $a_{\rm sl}^s$ is constrained by $B$-decay data 
to be too small to be accessible in measurements of decay rate asymmetries. On the other hand, 
we may extract $A_{\rm P}(B_s)$ and $a_{\rm CP}(\ell^+\nu_\ell;f_{D_s})$ from 
\begin{align}
A_{\rm P}(B_s) &= \frac{1}{2} \left(a_{\rm WS} + a_{\rm RS}-a_{\rm sl}^s\right), \\
a_{\rm CP}(\ell^+\nu_\ell;f_{D_s})  &= \frac{1}{2}\left(a_{\rm WS} - a_{\rm RS}-a_{\rm sl}^s\right),
\end{align}
where we have neglected the $F_\pm(t)$ terms and $a_{\rm sl}^s$ is constrained by
(\ref{asl-num-1}), playing a negligible role. This analysis opens a new way to determine both 
the CP asymmetry in $\bar B_s^0$ decays and the production asymmetry $A_{\rm P}(B_s)$.

From the experimental point of view, it is interesting to consider the following untagged 
rate asymmetry \cite{DFN}:
\begin{displaymath}
\hspace*{-1.4truecm}a_{\rm  unt}(t)\equiv
\frac{\Gamma[D_s^-\ell^+\nu_\ell,t]-\Gamma[D_s^+\ell^-\bar\nu_\ell,t]}{\Gamma[D_s^-\ell^+\nu_\ell,t]+
\Gamma[D_s^+\ell^-\bar\nu_\ell,t]}
\end{displaymath}
\vspace*{-0.5truecm}
\begin{displaymath}
=a_{\rm CP}(\ell^+\nu_\ell;f_{D_s})+\frac{a_{\rm sl}^s}{2}-\left[ \frac{a_{\rm sl}^s+2A_{\rm P}(B_s)}{2} 
\right]\hspace*{-0.1truecm}\left[\frac{\cos(\Delta M_st)}{\cosh(\Delta\Gamma_st/2)}\right]
\end{displaymath}
\vspace*{-0.5truecm}
\begin{equation}\label{untagged-asym}
+\frac{1}{2}\Delta{\cal A}_{\Delta\Gamma}\tanh(\Delta\Gamma_st/2),
\end{equation}
where $\Gamma[f,t]\equiv \Gamma(B^0_s(t)\to f)+\Gamma(\bar B^0_s(t)\to f)$. The LHCb 
collaboration employed (\ref{untagged-asym}) for the time-integrated untagged rates to determine 
(\ref{LHCb}). The term involving the production asymmetry is then essentially washed out due to the 
rapid $B^0_s$--$\bar B^0_s$ oscillations \cite{LHCb-2014}. However, the time dependence of 
(\ref{untagged-asym}) allows also the extraction  of $A_{\rm P}(B_s)$ and 
$a_{\rm CP}(B_s;f_{D_s})$, complementing the determinations proposed above. 

We have presented our new strategy for semileptonic decays. However, it actually applies to any 
flavor-specific $B^0_s$ mode, in particular $B^0_s\to D_s^-\pi^+$. Moreover, it can be applied to 
any $D^-_s\to f_{D_s}$ decay which is experimentally accessible. In contrast to conventional 
analyses of CP violation in such transitions, the new strategy is not affected by the production 
asymmetry 
\begin{equation}\label{APDs}
A_{\rm P}(D_s)=(-0.33 \pm 0.22 \pm 0.10) \%
\end{equation}
between the $D_s^+$ and $D_s^-$ mesons \cite{LHCb-Ds}. We advocate to implement 
the new method at LHCb and future runs of Belle II at the $\Upsilon(5S)$ resonance.

\section{Direct CP Violation}\label{sec:directCPV}
The key point of our new strategy is that the $D^-_s$ mesons, which are produced in the
$\bar B^0_s\to D_s^-\ell^+\nu_\ell$ transitions, will decay further as $D_s^-\to f_{D_s}$. 
Therefore, the rate asymmetries are sensitive 
to CP violation in both the $\bar B^0_s$ and the subsequent $D_s^-$ decays. 
Keeping only leading-order terms in CP-violating effects in the CP asymmetry 
defined in (\ref{acp-def}), we obtain
\begin{equation}\label{ACP-decomp}
a_{\rm CP}(\ell^+\nu_\ell;f_{D_s})=a_{\rm CP}^{(B_s)}|_{\ell^+\nu_\ell}+a_{\rm CP}^{(D_s)}|_{f_{D_s}},
\end{equation}
where 
\begin{equation}\label{acpBs-def}
a_{\rm CP}^{(B_s)}|_{\ell^+\nu_\ell}\equiv \frac{\Gamma(B^0_s\to D_s^-\ell^+\nu_\ell)-
\Gamma(\bar B^0_s\to D_s^+\ell^-\bar\nu_\ell)}{\Gamma(B^0_s\to D_s^-\ell^+\nu_\ell)+
\Gamma(\bar B^0_s\to D_s^+\ell^-\bar\nu_\ell)}
\end{equation}
probes CP violation at the $B^0_s$ amplitude level, whereas 
\begin{equation}
a_{\rm CP}^{(D_s)}|_{f_{D_s}}
\equiv \frac{\Gamma(D_s^-\to f_{D_s})-\Gamma(D_s^+\to \bar f_{D_s})}{\Gamma(D_s^-\to f_{D_s})+
\Gamma(D_s^+\to \bar f_{D_s})} 
\end{equation}
measures CP violation in the $D_s^-\to f_{D_s}$ processes.

Such ``direct" CP asymmetries can be generated through the interference between at least two 
decay amplitudes with non-trivial CP-conserving and CP-violating phase differences \cite{RF-rev}. 
The CP-conserving phases can be induced through strong interactions or absorptive parts of loop 
diagrams, while the CP-violating phases are provided by the CKM matrix in the SM or NP effects.

In the SM, $a_{\rm CP}^{(B_s)}|_{\ell^+\nu_\ell}$ is zero at leading order in weak interactions and takes a 
vanishingly small value through higher-order effects \cite{BSEGR,JZ,BZ}. Even in the presence of 
NP, this CP asymmetry cannot take sizeable values. For the non-leptonic decay $B^0_s\to D_s^-\pi^+$ 
things have to be assessed more carefully, as there may still be room for NP at the decay amplitude 
level \cite{BLTXW,JKLTX} and strong interactions are at work. It would be interesting to measure 
\begin{equation}
a_{\rm CP}(\pi^+;f_{D_s})-a_{\rm CP}(\ell^+\nu_\ell;f_{D_s})=a_{\rm CP}^{(B_s)}|_{\pi^+}
\end{equation}
with our method, where $a_{\rm CP}^{(D_s)}|_{f_{D_s}}$ cancels. 

Concerning direct CP violation in $D_s$ decays, non-leptonic channels play the key role. In the SM, 
the CP asymmetries are small but may be enhanced through NP \cite{BFBB,MG}. Predictions suffer 
from hadronic uncertainties, where the $SU(3)$ flavor symmetry offers a useful tool \cite{MNS}. 

LHCb employed $D_s^\mp\to K^+K^-\pi^\mp$ modes for the analysis of $a_{\rm sl}^s$. Consequently,
the experimental result in (\ref{LHCb}) actually probes 
$(a_{\rm sl}^s)_{\rm eff}=a_{\rm sl}^s + 2 a_{\rm CP}^{(D_s)}|_{K^+K^-\pi^\mp}$.
Using the constraint for $a_{\rm sl}^s$ in (\ref{asl-num-1}), we may convert (\ref{LHCb}) into
\begin{equation}\label{Ds-LHCb}
a_{\rm CP}^{(D_s)}|_{K^+K^-\pi^\mp}=\left(0.19 \pm 0.17\right)\times10^{-2}, 
\end{equation}
which is five times more precise than the average 
$(0.5\pm0.9)\times 10^{-2}$ \cite{PDG} of CLEO measurements \cite{CLEO1,CLEO2} and
about two times more precise than (\ref{APDs}), thereby illustrating the potential of the new method.

To probe NP, decays with penguin loop contributions, such as 
$D_s^+\to\pi^+K^0, \pi^0K^+, K^+\phi$, are most promising. Taking $D_s^+\to\pi^+K_{\rm S}$
as an example \cite{LX,BGR}, LHCb measured 
$a_{\rm CP}^{(D_s)}|_{K_{\rm S}\pi^\pm }=-(0.38\pm0.46\pm0.17)\%$ \cite{LHCb-DspiKS}, 
which has an uncertainty three times larger than (\ref{Ds-LHCb}).

\section{Conclusions}
We are moving towards impressive new frontiers in high-precision studies of CP violation. The global 
agreement of the current data with the Standard Model raises the question of how much space 
is left for New Physics. In the case of CP violation in $B^0_s$--$\bar B^0_s$ oscillations, we have used 
experimental data to obtain $a_{\rm sl}^s = (0.004\pm 0.075)\times10^{-2}$, pushing this observable
well below the currently accessible regime. We have presented a model-independent formalism in terms of 
CP-violating phases and decay rates, allowing us to further refine this range through improved data 
for CP violation in decays of the kind $B^0_s\to J/\psi \phi$ and $B^0_s\to D_s^{(*)+}D_s^{(*)-}$. 

The $a_{\rm sl}^s$ observable has been in the focus for many years and was determined by means of 
flavour-specific $B^0_s\to D_s^-\ell^+\nu_\ell$ decays. In view of our findings for $a_{\rm sl}^s$, we 
propose a new strategy to utilize such decays. It allows us to determine the $B^0_s$ production 
asymmetry $A_{\rm P}(B_s)$ and the CP asymmetry $a_{\rm CP}(\ell^+\nu_\ell;f_{D_s})$. The current 
experimental situation for $A_{\rm P}(B_s)$ leaves room for significant improvement, while 
$a_{\rm CP}(\ell^+\nu_\ell;f_{D_s})$ is governed by the direct CP violation in $D_s^-\to f_{D_s}$, thereby 
opening a new avenue for the exploration of this phenomenon. 

We have also given expressions allowing us to probe whether $B^0_s\to D_s^-\ell^+\nu_\ell$ decays 
are actually flavor-specific. This feature, which is commonly used, is most plausible but has not yet 
been tested experimentally. 

The new strategy can also be applied to other flavor-specific modes, such as $B^0_s \to D_s^- \pi^+$. 
It shifts flavor-specific $B^0_s$ decays from their well-known role as probes of CP violation in 
$B^0_s$--$\bar B^0_s$ oscillations to new tools for the exploration of direct CP violation, in particular 
in $D_s^\pm$ decays. As CP violation in these modes might be enhanced by New Physics, we have a 
new framework to search for such signatures, allowing us to take full advantage of the corresponding 
physics potential in the high-precision era of heavy-flavor studies.

\vspace*{0.3truecm}

\section*{Acknowledgements} 
We would like to thank Marcel Merk and Vincenzo Vagnoni for useful discussions.
This work is supported by the Foundation for Fundamental Research on Matter (FOM) and 
by the Deutsche Forschungsgemeinschaft (DFG) within research unit FOR 1873 (QFET).

%
%
%

%
%
%
\end{document}